\documentclass[11pt]{article}

\usepackage{arxiv}
\usepackage{latexsym,subfigure,epsfig}
\usepackage{amsfonts}
\usepackage{amsmath,paralist, amssymb}
\usepackage{times}
\usepackage{graphics,color}
\usepackage{url}
\usepackage{graphicx}
\usepackage{wrapfig,enumitem}

\usepackage[utf8]{inputenc} 
\usepackage[T1]{fontenc}    

\usepackage{booktabs}       
\usepackage{nicefrac}       
\usepackage{microtype}      
\usepackage{lipsum}

\newtheorem{insight}{Insight}
\newcommand{\ignore}[1]{}

\usepackage[square,numbers]{natbib}
\ignore{
\usepackage{fancyhdr}

\pagestyle{fancy} 

\fancyhf{}

\lhead{Approved for Public Release; Distribution Unlimited. MITRE Public Release Case Number 21-2666}
\rhead{}
}
\begin{document}

\title{Social Engineering Attacks and Defenses in the Physical World vs. Cyberspace: A Contrast Study}

\author{Rosana Monta\~nez Rodriguez$^1$, Adham Atyabi$^2$, and Shouhuai Xu$^2$ \\
$^1$Department of Computer Science, University of Texas at San Antonio \\ San Antonio, TX, USA 78249 \\
$^2$Department of Computer Science, University of Colorado Colorado Springs \\ Colorado Springs, CO, USA 80918 
}

\date{}

\maketitle

\begin{abstract}
Social engineering attacks are phenomena that are equally applicable to both the physical world and cyberspace. These attacks in the physical world have been studied for a much longer time than their counterpart in cyberspace. This motivates us to investigate how social engineering attacks in the physical world and cyberspace relate to each other, including their common characteristics and unique features. For this purpose, we propose a methodology to unify social engineering attacks and defenses in the physical world and cyberspace into a single framework, including: (i) a systematic model based on psychological principles for describing these attacks; (ii) a systematization of these attacks; and (iii) a systematization of defenses against them. 
Our study leads to several insights, which shed light on future research directions towards adequately defending against social engineering attacks in cyberspace.
\end{abstract}

{\bf Keywords}: Social Engineering Attacks, Social Engineering Defenses, Cybersecurity, Human Cognition, Human Factors, Phishing.

\section{Introduction}

Social engineering attacks are prevalent in both the physical world and cyberspace. Intuitively, these attacks attempt to cause an error, or failure, in a target or victim's decision-making process to benefit the attacker. The prevalence of these attacks can be attributed to their low cost and effectiveness. In the physical world, social engineering attacks share many similarities with scams and fraud. In cyberspace, social engineering attacks are often the first step of sophisticated attacks that can cause substantial damages.

\subsection{Our Contributions}

In this chapter, we make four contributions. First, we propose a methodology to unify social engineering attacks in the physical world and their counterpart in cyberspace into a single framework. The methodology is novel because it takes a unique perspective based on the following observation.
In principle, a social engineering attack attempts to manipulate a victim into complying with a request from the attacker by leveraging aspects of {\em individual and social cognition}, which provides a compelling perspective for the study. Individual cognition examines the internal processes that lead to decision-making and behavior in an individual, whereas social cognition explores the external social aspects that affect these internal processes. These two complementary aspects of cognition provide a basis for interpreting information: one based on a self-centered perspective influenced by mental processing of sensory input, and the other influenced by the interaction with other humans. This perspective guides us to propose a model for describing attacker-victim interactions in both the physical world and cyberspace. 

Second, we systematize social engineering attack techniques in the physical world and cyberspace. In total, we systematize seven techniques (belonging to five categories) used in social engineering attacks in the physical world and 13 techniques (belonging to four categories) used in social engineering attacks in cyberspace. To the best of our knowledge, this is the first systematization of social engineering attack techniques in the physical world and cyberspace.

Third, we systematize defenses against social engineering attacks (i.e., social engineering defenses) in the physical world and cyberspace. For social engineering defenses in the physical world, we systematize six defenses (belonging to two categories, namely preventive defenses and proactive defenses).  For social engineering defenses in cyberspace, we systematize 11 defenses (belonging to three categories, namely preventive defenses, proactive defenses, and reactive defenses). To the best of our knowledge, this is the first systematization of social engineering defenses in the physical world and cyberspace.

Fourth, we conduct a contrast analysis of the social engineering attacks and defenses in the physical world and cyberspace. The analysis draws a number of insights, such as: (i) we should strive to achieve social engineering resistance by design; (ii) there are no ``silver bullet'' defenses that can work against both social engineering attacks in the physical world and cyberspace because these two worlds exhibit different features and demand tailored solutions. The analysis sheds light on future research. 

\subsection{Related Work}
\label{sec:related-work}

We reviewed prior studies on social engineering attacks through the lens of cognition, leading to the distinction of {\em individual} vs. {\em social cognition}. The former studies the mental processes that affect attention, memory, perception, decision-making, and information processing; the latter studies how social interactions affect those mental processes. These prior studies are relevant to the present one because they provide insight into the factors and interactions that facilitate the exploitation of human psychological weaknesses in social engineering attacks. 
Since most studies on social engineering attacks in the physical world are based on anecdotal accounts \cite{stajano2009cambridge, dove2020psychology}, meaning there is a lack of peer-reviewed, quantitative experiments,
we mitigate this limitation by including studies in scams, because scams share many characteristics with social engineering attacks.

\subsubsection{Prior Studies Related to Individual Cognition}
We adopt the distinction between cognitive \emph{attributes} and \emph{attitudes} described in the literature, namely that an \emph{attribute} is ``an emergent property of an individual embodied in social practices" \cite{guyon2017modeling} whereas an \emph{attitude} is the ``relatively enduring predisposition to respond favorably or unfavorably toward something" \cite{simon1976adm}.  

\smallskip

\noindent{\bf Prior Studies Related to Individual Cognitive Attributes}. We consider the following attributes that are related to social engineering attacks: {\em personality}, {\em expertise}, {\em individual difference}, {\em culture}, {\em workload}, {\em stress}, and {\em vigilance}. These attributes affect one's susceptibility to social engineering attacks.

For \emph{personality}, there is no consensus on its relationship with one's susceptibility to social engineering attacks in cyberspace.\ignore{\footnote{be precise; if some studies are in one world and others are in the other, specifically state which paper is conducted in which world---physical vs cyber}}
Personality can be characterized by five main domains, namely {\em neuroticism}, {\em openness}, {\em extroversion}, {\em consciousness},\ignore{\footnote{is this personality?}} and {\em agreeableness}. We revise the state-of-the-art understanding about their relationships to social engineering attacks.
(i) In terms of neuroticism, two studies suggest that a high neuroticism is associated with lower self-efficacy (i.e., user confidence to manage a cyber risk) \cite{halevi2016cultural} and increase one's susceptibility to phishing \cite{halevi2013pilot}, but a study on phishing \cite{cho2016ieee} suggests that a high neuroticism decreases one's susceptibility to phishing attacks. 
(ii) In terms of {\em openness}, one study \cite{halevi2013pilot} suggests that a high openness increases one's susceptibility to privacy attacks, but other studies \cite{halevi2016cultural,pattinson2012imcs} suggest a high openness reduces one's susceptibility to phishing attacks.
(iii) In terms of {\em extroversion}, one study \cite{lawson2018iergoa} suggests that a high extroversion increases one's susceptibility to phishing attacks, 
but another study \cite{pattinson2012imcs} suggests that a high extroversion decreases one's susceptibility to phishing attacks.
(iv) In terms of {\em consciousness}, two studies \cite{halevi2016cultural,lawson2018iergoa} suggest that a high consciousness reduces one's susceptibility to phishing, but another study \cite{halevi2015ssr} suggests that a high consciousness increases one's susceptibility to targeted social engineering attacks.
(v) In terms of {\em agreeableness}, two studies \cite{cho2016ieee,darwish2012ieee} show that a high agreeableness increases one's susceptibility to phishing attacks. 

The impact of the rest of the factors are briefly reviewed as follows.
For \emph{expertise}, cybersecurity expertise does reduce one's  susceptibility to  phishing \cite{kumaraguru2006acm}. In the physical world, it improves threat appraisal and risk perception \cite{klein1991ieee_smc} 
For non-experts, experience in combination with knowledge reduces one's susceptibility to phishing
\cite{harrison2016Oinforev, abbasi2016ieee, gavett2017plos, pattinson2012imcs,wright2010mis, downs2006acm}, and malicious social media messages \cite{redmiles2018acm}. 
But pure awareness of cyber threats does not appear to 
reduce one's susceptibility to social engineering attacks \cite{junger2017priming, sheng2010falls,downs2006acm}. However, in scams, knowledge of the threat reduces vulnerabilities \cite{langenderfer2001consumer}, but domain specific knowledge may increase susceptibility to scams \cite{lea2009psychology}.
For \emph{individual differences},  
age increases both young 
(18-25) \cite{sheng2010falls, howe2012psychology} and old (65+) \cite{howe2012psychology, gavett2017plos} people's susceptibility to phishing and online threats. Additionally, older people are generally more susceptible to spear-phishing \cite{lin2019acm} and scams both in the physical world \cite{langenderfer2001consumer} and cyberspace \cite{grimes2007email}. 
Most studies have found no relationship between one's gender and susceptibility to social engineering attacks in cyberspace \cite{sawyer2018humanfactors, purkait2014iscs, rocha2014imcs,bullee2017ics} or in the physical world \cite{lea2009psychology}. 
For \emph{culture}, studies show that individuals are more susceptible to social engineering messages \cite{alhamar2010ieee, bohm2011mt, Sharevski2019arXirv,tembe2014acm,redmiles2018acm} and scams \cite{dove2020psychology} that align with their cultural norms.  
For \emph{workload}, it is known that 
a high email load \cite{vishwanath2011_dssystems},
work overload \cite{jalali2020jmir}, and inattentional blindness as a result of workload \cite{pfleeger2012leveraging},  might increase one's susceptibility to social engineering attacks.
For \emph{stress}, one study \cite{stajano2009cambridge} shows that stress increases one's susceptibility to scams attacks.
For \emph{vigilance}, one study \cite{purkait2014iscs} shows that a high attentional vigilance reduces one's susceptibility to phishing websites. In the physical world, vigilance reduces scam victimization \cite{dove2020psychology}.  

\smallskip

\noindent{\bf Prior Work Related to Individual Cognitive Attitudes}. We consider the following attitudes that are related to social engineering attacks: {\em trust attitude}, {\em suspicion attitude}, 
and {\em risk attitude}. These attitudes affect one's susceptibility to social engineering attacks.
For \emph{trust attitude}, a high trust attitude incurs a high susceptibility to social engineering attacks \cite{workman2008wisecrackers, abbasi2016ieee, halevi2013pilot,rocha2014imcs} and scams \cite{langenderfer2001consumer}.
For \emph{suspicion attitude}, an individual with a high suspicion is less susceptible to social engineering attacks \cite{vishwanath2018commresearch, tembe2014acm,wright2010mis}. 
For \emph{risk attitude}, a high risk perception reduces one's susceptibility to social engineering attacks  \cite{halevi2015ssr,rocha2014imcs,halevi2016cultural,howe2012psychology,sheng2010falls} and low risk perception increase susceptibility to scams \cite{dove2020psychology}. 


\subsubsection{Prior Work Related to Social Cognition}

Two factors have been investigated in the literature.  One is \emph{persuasion}, where social engineering attacks \cite{van2019cognitive,rajin2018frontiers} and scams \cite{lea2009psychology,langenderfer2001consumer} with a high persuasive capability are more successful.
The other is \emph{scam compliance} where using persuasion along with emotional and visceral triggers increases scam successes \cite{stajano2009cambridge, lea2009psychology}.


\subsection{Outline}
The rest of the chapter is organized as follows. Section \ref{sec:methdology} presents the terminology and methodology used in this study. Section \ref{sec:physical-world} characterizes social engineering attack model, techniques ans defenses in the physical world. Section \ref{sec:cyberspace} characterizes social engineering attack model, techniques ans defenses in the physical world. Section \ref{sec:contrast-analysis} describes our contrast analysis between the physical world and cyberspace and future research directions. Section \ref{sec:conclusion} concludes the present chapter.

\section{Terminology and Methodology}
\label{sec:methdology}

\subsection{Terminology}


We propose using the following terminology to describe social engineering attacks in the physical world and cyberspace, while noting that many of these terms are adapted from cyber attacks terminology. The term {\em target}, which is often used in cyber social engineering literature, describes a human being who may have some exploitable psychological weakness which can be leveraged by the attacker in question. 
Since the term {\em victim} is often used in other social engineering literature, we will use these two terms interchangeably. 
A {\em target} (i.e., victim) can be characterized by an {\em attack surface}, which is defined as the set of vulnerabilities (i.e., attack vectors) that can be exploited by the attacker to victimize the  target. We use the term {\em social engineering attacks} to describe the attacks that exploit psychological weaknesses or vulnerabilities of humans to achieve malicious goals, such as acquiring information, access or assets.


    \ignore{"for the purpose of disrupting, disabling, destroying, or maliciously controlling a computing environment/infrastructure; or destroying the integrity of the data or stealing controlled information."\cite{CyberAtt62:online}
    }

\subsection{Methodology}

Our methodology is inspired by the following observations. Information processing is central to understanding social engineering attacks because (i) it describes the process by which external sensory inputs are processed by internal cognitive units to interpret information and (ii) its output shapes one's decision making, judgement and behavior. Social engineering attacks involve the processing of persuasive messaging to influence a desired behavior. The Elaboration Likelihood Model (ELM) has been used to study information processing of persuasive messages  \cite{cacioppo1984elaboration}. ELM is a dual information processing system in which the processing of information can take one of two routes: {\em central} vs. {\em peripheral} route, despite that the information processing path can alter between these two routes in the course of information processing. Peripheral processing is fast, based on heuristics and requires less cognitive effort; whereas central processing is slow, analytical and requires cognitive effort. Studies have demonstrated how the processing route affects the outcome of a social engineering attack. A social engineering attack often succeeds by making a target trap into the peripheral processing route \cite{lea2009psychology, dove2020psychology}. That is, a sophisticated social engineering attack often induces peripheral information processing and an effective defense should trigger central information processing. Although factors beyond the attacker's control  may also affect the information processing route (e.g., the environment where the interaction occurs), in this paper we focus on the factors that are under the attacker's control. Two key factors that affect the selection of routes are {\em trust} and {\em suspicion}, where trust encourages peripheral processing and suspicion encourages central processing. 

Although social engineering attacks in cyberspace exploit the same psychological principles that are exploited by social engineering attacks in the physical world, there is a significant difference between social engineering attacks in these two worlds, namely how communication is mediated. This difference may affect victims' performance and provide different opportunities to the attackers. The proceeding observations prompt us to characterize social engineering attacks in the two worlds via: {\em model}, {\em attacks} and {\em defenses}.

\smallskip

\begin{figure*}[!htpb]
\centering
\includegraphics[width=\textwidth,keepaspectratio]{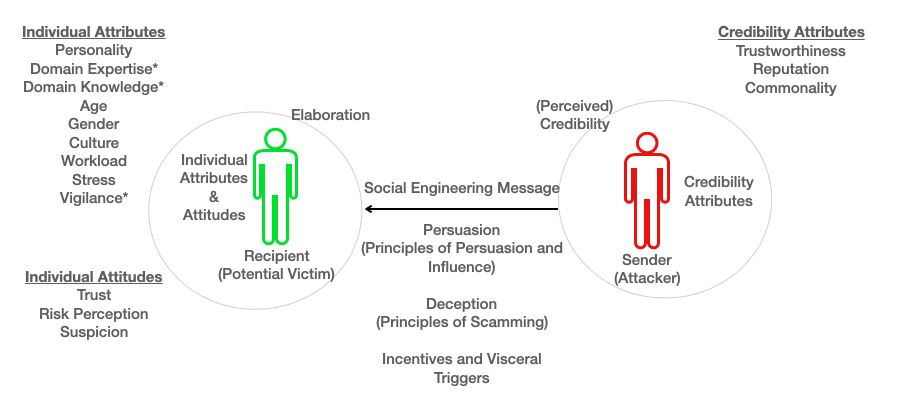}
\caption{Our social engineering attack model, where 
asterisk (*) indicates that the attribute in question can reduce one's susceptibility to social engineering attacks.
}
\label{fig:model-in-physical-world}
\end{figure*}

\noindent{\bf Social Engineering Attack Model}. Figure \ref{fig:model-in-physical-world} highlights our social engineering model, which illustrates the elements and interactions pertinent to social engineering attacks. The model describes the three main components of a social engineering attack: \emph{the attacker}, \emph{the message}, and \emph{the victim}. The model is equally applicable to  social engineering attacks in the physical world and in cyberspace.

\ignore{We start our analysis of social engineering by presenting a model that describes the communication elements and psychological attributes that affect the interaction between recipient and attacker (Figure \ref{fig:model-in-physical-world}). The social engineering model applies to both social engineering attacks in the physical world and cyberspace. In section \ref{sec:model_se-cy}, we then present distinct environmental features that affect this interaction in cyberspace.}

\smallskip

\noindent{\bf Social Engineering Attack Techniques}. In order to describe social engineering attacks, we use the term {\em attack kill chain} to describe the sequence of orchestrated attack {\em phases}, where each {\em attack phase} can be described by a high-level abstraction that provides a logical grouping of distinct actions conducted by the attacker. From the planning to the termination of an attack, there are multiple phases, such as those specified by the the Lockheed Martin kill chain \cite{CyberKillChainPaper2011} and the Mandiant kill chain \cite{Mandiant}. {\em Phases} are divided into {\em tactics} and {\em techniques}, which are adapted from the MITRE ATT\&CK framework \cite{ATTCK10170:online}. The term {\em tactic} describes a short-term objective of performing an attack action and the term {\em technique} describes the actions performed in support of a tactic \cite{ATTCK10170:online}. Although techniques are independent activities, they often supplement or assist one another. For example, during the {\em planning phase}, an attacker would execute a  {\em reconnaissance tactic} to gather information about a target and identify vulnerabilities and methods for exploiting these vulnerabilities. In support of this tactic, the attacker can use some of the following techniques: {\em passive surveillance}, {\em dumpster diving}, and {\em open source reconnaissance}.

\smallskip

\noindent{\bf Social Engineering Defenses}. In order to defend against social engineering attacks, multiple kinds of defenses can be employed. The following terms are adapted from their counterparts in cyber defense context  \cite{XuCybersecurityDynamicsHotSoS2014,XuBookChapterCD2019,XuMTD2020,XuTAAS2012,XuTAAS2014,XuInformationSystemsFrontiersEditorial2021,XuIEEETNSE2018,XuIEEEACMToN2019,XuTNSE2021-GlobalAttractivity,Pendleton16}: {\em preventive defenses} aim to prevent a target from falling victim to social engineering attacks; {\em proactive defenses} aim to mitigate the attacks that may have been successful but are not detected by (or known to) the target and/or the defender; {\em reactive defenses} aim to detect successful attacks and recover from the compromised state to the secure state. Note that reactive defenses do not apply to the physical world because the targets are humans who would be punished once detected; whereas in cyberspace, the victims are often innocent and their compromised computers need to recover from a compromised state to a secure state.

\section{Characterizing Social Engineering Attack Model, Techniques and Defenses in the Physical World}
\label{sec:physical-world}

In the physical world, social engineering attacks are characterized by  face-to-face interactions between an attacker and a victim. On one hand, face-to-face interactions increase the victim's trust in the attacker \cite{riegelsberger2003researcher} and  allow the attacker to tailor attacks against the victim \cite{alexander2016sans}. On the other hand, face-to-face interactions often expose attacker identities, increasing the risk of exposure \cite{dimkov2010csac}.

\subsection{Social Engineering Attack Model of the Physical World} \label{sec:model_se-phy}

As highlighted in Figure \ref{fig:model-in-physical-world}, we consider the following social engineering model of attacks against victims. 

\subsubsection{Attacker}

An attacker's goal is to manipulate a victim into performing an action that would grant the attacker access to the intended information or asset. An attacker's weapon to manipulate the victim into compliance is through message exchange. 
To be effective, an attacker must keep victims' risk perceptions low by projecting qualities associated with credibility and approaching a victim with a message or offer which induces peripheral processing.  Credibility is a victim's perception that the attacker will deliver on the offer. Credibility increases trust and lowers risk perceptions \cite{stajano2009cambridge}. An attacker's credibility is characterized by the following attributes: {\em commonality}, {\em reputation}, and {\em trustworthiness}.

\ignore{social engineers also leverage external factors that increase an individual's cooperation. Factors that influence cooperation are 1) time, 2) institution and 3) social network \cite{raub2000management}. Time defines the likelihood of future interactions [expand]. }
    
Commonality is the perceived common ground between a victim and an attacker. By establishing commonality, an attacker can inherit the trust extended to members of the group. Commonality can be established by providing the details that are only known to group members (contextualization), demonstrating familiarity with the victim (personalization), or sharing common biases and beliefs.   \ignore{\cite{guadagno2013social} Message Personalization and Contextualization.     Relationship between bias and belief, and trustworthiness and incomplete information on peripheral processing (personalization \cite{hirsh2012psysci})}

Reputation is a property describing the assessment of others about an individual or source. Reputation is often extrapolated based on characteristics like associates (or social network) and affiliation to institutions. Reputation increases cooperation, explaining why social engineering attacks often exploit both social networks and affiliations to reputable institutions. For example, an attacker might assume the persona of an authority (e.g., government agencies like Internal Revenue Services, or law enforcement) or imply common social network connections with a victim to elicit cooperation.
    
Trustworthiness is related to trust, which is an individual's ``willingness to be vulnerable based on positive expectations about the actions of others" \cite{adams1999user}. Projecting trustworthiness requires an attacker to be perceived as providing services in good faith. Trust can also be developed through continuous interactions between a victim and an attacker. This kind of trust, known as {\em affection trust} \cite{mcallister1995affect}, is based on an affection bond built over time.  Under this condition, an individual might willingly take risks based on the relationship and disregard their risk perceptions.

\subsubsection{Message}
In order to be successful, an attacker's message would intend to induce peripheral processing. For messages involving a high-risk request, an attacker might wage multiple interactions with a victim to achieve compliance because multiple message exchanges might allow an attacker to develop a familiar relationship with a victim, increasing the levels of trust \cite{mcallister1995affect} and making the risk more acceptable for the victim.  
To increase the chance of success, the following psychological techniques could be leverage to craft messages: {\em persuasion}, {\em scamming}, {\em incentive and motivator}, and {\em visceral trigger}.
(i) Persuasion is the act of presenting an argument that encourages an individual to behave in a desired manner \cite{cialdini2007influence}.
(ii) Scamming (or deception) is the act of presenting an argument with the intention to create a false belief \cite{buller1996idt,stajano2009cambridge}.
(iii) Incentive and motivators encourage cooperation \cite{lea2009psychology,dove2020psychology}, where incentives leverage external rewards and motivators leverage internal psychological attributes.
\ignore{An example of incentives is quid pro quo. Many users that fall for social engineering attacks ignore the risk of their actions and focus on the benefits or potential benefit of the phishing email \cite{halevi2015ssr, halevi2013pilot}.   Users will also trade privacy for convenience, or bargain release of information for a reward \cite{workman2007iss}.}
(iv) Visceral triggers are motivational manipulations which trigger emotional response by exploiting needs and desires \cite{stajano2009cambridge, lea2009psychology}.

\subsubsection{Victim}

A victim's goal is to identify social engineering attacks while avoiding a high false-positive rate. A social engineering attack succeeds when the victim complies with the attacker's request. In the physical world, the attacker might have to ensure that the victim feels positive about their interaction after that the victim complies, which reduces the regret that discourages reporting of the incident to authorities. 

To prevent victimization, the recipient of a message must process it through the central route. Activating central processing requires that the victim detects inconsistencies and deception cues in the message. Attributes facilitating this include {\em domain expertise}, {\em domain knowledge}, and {\em vigilance}.
(i) Domain expertise can reduce victimization to social engineering attacks in several ways:  individuals with expertise have more accurate threat mental models which improve threat appraisal and risk perceptions \cite{klein1991ieee_smc}, while having better strategies to cope with threat when their risk assessment is erroneous; domain expertise also facilitates the detection of deceptive cues. 
(ii) Domain knowledge can be developed through training or previous negative experience. It helps with pattern recognition and deceptive cue detection \cite{langenderfer2001consumer}, while noting that deceptive cue detection is a precursor to suspicion.  (iii) Vigilance is the process of dedicating cognitive resources to perform a demanding task, such as detecting cues which can indicate deceptive intent in a message \cite{dove2020psychology, duffield2001psychology}. Note that vigilance is affected by suspicion.

\subsection{Social Engineering Attack Techniques in the Physical World}
\label{sec:phy-techniques}

\begin{figure*}[!htpb]
\centering
\includegraphics[width=\textwidth,keepaspectratio]{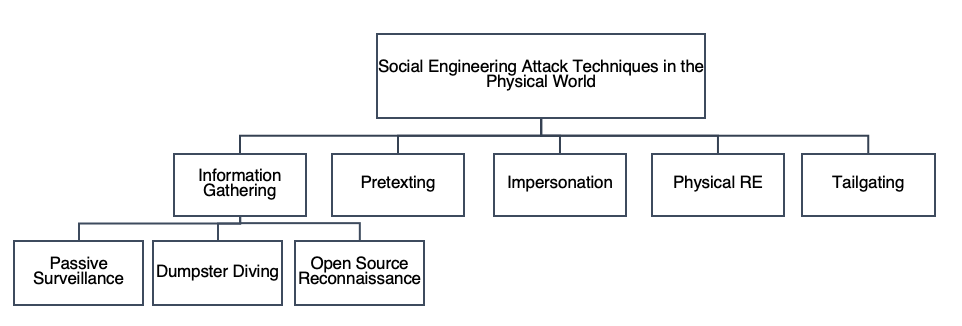}
\caption{Taxonomy of social engineering attack techniques in the physical world.
}
\label{fig:taxonomy-ph}
\end{figure*}

As highlighted in Figure \ref{fig:taxonomy-ph}, we classify social engineering attack techniques in the physical world into five categories: {\em information gathering}, {\em pretexting}, {\em impersonation}, {\em physical reverse engineering} (physical RE), and {\em tailgating}, which are elaborated below.


\subsubsection{Information Gathering}
This category of attack techniques focuses on acquiring information about a target. These techniques can support social engineering in the physical world. This category has three specific attack techniques: {\em passive surveillance}, {\em dumpster diving}, {\em open source reconnaissance}.
First, the passive surveillance attack technique \cite{greenlees2009iee} attempts to collect information about a victim and the environment. The information is used in later phases of an attack to develop a cover story and artifacts supporting an objective without raising the victim's suspicion. 
Second, the dumpster diving attack technique is the act of searching through the trash for information \cite{mitnick2003art}. It is effective because all the content in the trash is specific to the victim. The information can be used along with other social engineering attack techniques \cite{redmon2005mitigation}. 
Third, the open source reconnaissance attack technique is the gathering of information that is publicly accessible \cite{mitnick2003art}. In the past, libraries were one of the main sources of gathering open-source information \cite{thompson2006helping}. Nowadays, this attack often involves the mining of information available online and in social media \cite{ariu2017social}. For example, a simple google search on the name of a person of interest would result in a set of links, images, and videos that either directly involves the person or have been searched/accessed by the person. These types of information can be used to generate a psycho-behavioral profile for the target person to generate a personalized social engineering attack.

\subsubsection{Pretexting}

The pretexting attack technique attempts to obtain information by using false pretenses \cite{baer2008corporate}. It requires that the attacker invents a background story to create a scenario that is relevant to the victim and persuade the victim to perform an action or release information \cite{indrajit2017social}. For example, in the HP pretexting scandal {\cite{baer2008corporate}}, the HP security department hired a third-party investigator to identify the source of private HP Board conversations disclosure to the press. Using pretexting, the third-party investigator was able to acquire records from phone service providers by impersonating HP Board members and journalists that were suspected to be involved in the leak \cite{workman2008wisecrackers}.  

\subsubsection{Impersonation}
The impersonation attack technique uses a persona (e.g., that can increase compliance of a victim. Personas allow the attacker to keep a low profile and blend into the targeted environment. Examples of personas are authority personas \cite{greenlees2009iee} (e.g., manager or IT auditor), or  a layperson persona \cite{redmon2005mitigation} (e.g., custodian or delivery person). Persona can  facilitate access to information, assets or places. Personas also allows the attacker to leverage different persuasion techniques. Authority persona allow an attacker to leverage the persona perceived position and potential consequences if the victim does not comply.  Persona selection can be based on the security of the environment were the asset is maintained, or based on the individual that is responsible for safeguarding the asset \cite{dimkov2010csac}. For example, an attacker might impersonate a company employee to convince a cleaning staff employee to give them access to an asset. An attacker that focuses on exploitation of the custodian of an asset might choose to impersonate service desk staff, a coordinator representative or an individual that needed urgent access to the asset. 
    
    \ignore{
    Focusing in the environment allow an attacker leverage the An environment Focus (EF) methodology test the security of the environment were the asset is maintained. Custodian Focused (CF) focuses on the individual that is responsible for safeguarding the asset. Similarly to EF, CF methodology relies on impersonation. For the CF penetration test, students impersonated a service desk staff, a coordinator representative or an individual that needed urgent access to the asset. 
    
    They presented the custodian with a fake email that “authorized” them to retrieve the asset. In 5 out of 10 tests, the students were able to gain access after presenting the email to the custodian. In three cases, the custodian reported the incident to a supervisor or campus security.  CF failed in 2 cases but the students were able to gain access to the asset using EF. Although the experiments were performed in a campus setting using university students as social engineers, they provide a good overview on items to consider when performing impersonation.}
    
\subsubsection{Physical Reverse Engineering (Physical RE)}
The physical reverse engineering attack technique requires that an attacker creates problem that gives them access to the victim and then offers assistance to fix it \cite{gragg2003multi}. This attack technique may proceed in the following three steps \cite{nelson2001methods}. (i) Sabotage: the attacker introduces a fault that causes a problem for the victim. (ii) Advertising: once the victim recognizes the problem, the attacker makes it known that the attacker can provide assistance which gives the attacker access to the target. (iii) Assisting: with the victim's consent, the attacker fixes the problem, while using the opportunity as a mechanism to launch an attack. 
An example of this attack technique in the  physical  world is an attacker modifying a system to give the appearance that the system is corrupted by displaying an error message. When the user notices that the system is "corrupted," they reach out to the attacker for help, because the attacker had previously advertised its expertise by leaving behind business cards or providing their contact information in the error message \cite{nelson2001methods}. 

 \subsubsection{Tailgating}
The tailgating attack technique involves gaining access into a control access facility or a restricted area \cite{alexander2016methods}, by following an individual with access into the facility. This technique is often combined with impersonation. In a typical scenario the attacker impersonates a delivery service individual. When an authorized individual opens the door, the attacker asks the individual to hold the door open or simply follows the individual without their notice.

\subsection{Social Engineering Defenses in the Physical World}
\label{sec:defense-phy}

\ignore{\footnote{\color{magenta}Dr Xu, I think would be a good approach for the unified framework \color{red} ideally, it'd be great to draw a picture that is in spirit similar to Figures 5, 14, 15 in this paper https://arxiv.org/pdf/1908.04507.pdf. when present the defenses, make connections to the structure that is used to characterize the attacks What are the requirements in order for a defense to be successful}}

It would be ideal that humans could identify or recognize social engineering attacks in the physical world. However, human cognitive resources are limited, while other tasks are also competing for the limited cognitive resources. For example, vigilance is cognitively costly and unsustainable over extended periods of time \cite{warm2018vigilance}. In order to address these limitations, social engineering defenses should follow a multi-layered defense approach (i.e., defense-in-depth). 
Corresponding to the methodology described above, we propose classifying social engineering defenses in the physical world into two categories: {\em preventive} defenses, which aim to prevent social engineering attacks from succeeding; 
{\em proactive} defenses, which aim to mitigate the attacks that may have been successful but are not detected by the target. {\em Reactive} defenses do not appear to be relevant here because humans are the target in the physical world, rather than computers.

\subsubsection{Preventive Defenses}  \label{sec:phy-prev}

There are five kinds of preventive defenses: {\em legislation}, {\em security controls}, {\em training}, {\em organizational policies}, and {\em organizational procedures}.

\smallskip

\noindent{\bf Legislation}. This approach focuses on deterring, disincentivizing or discouraging social engineering attacks, by increasing the personal risk to the attacker. For example, the HP pretexting scandal \cite{baer2008corporate} led to the United States Telephone Record and Privacy Act (2006), which criminalizes the employment of fraudulent tactics to persuade telephone companies to release phone records, with a punishment of up to ten years of prison. Prior to this Act, only pretexting for financial records was illegal under the Financial Services Modernization Act (1999). Similar to pretexting, impersonation (i.e., false identity)  is illegal under several laws in the United States, when it is used to cause harm or gain benefits. One of these laws is  18 U.S. Code § 912, which criminalizes the impersonation of an officer or servant of the United States government. Charges under this law can carry a maximum sentence of three years and/or a fine. It is difficult to quantify the impact of legislation on criminal activities, in part because legislation assists with the allocation of resources for prevention as well as increasing awareness of an issue which in turn reduces its incidence \cite{akirav2018model}. 

\smallskip

\noindent{\bf Access Controls}. One approach to defending against social engineering attacks like tailgating is to employ credential-based access in controlled areas \cite{redmon2005mitigation,cissp2009official, abeywardana2016layered}. This mechanism often requires individuals to scan their badge and/or enter their Personal Identification Number (PIN) to access a facility in question. Although this approach can be effective in preventing random individuals from tailgating, it does not stop all tailgating because it can be bypassed when an attacker is accompanied by an authorized individual \cite{cheh2019leveraging}. In order to prevent this attack, authorized individuals need guidance in dealing with situations where their risk perception level might be low.
    
\smallskip

\noindent{\bf Training}. Training (e.g., security education, awareness, resistance training) is a widely employed defense against social engineering attacks in the physical world \cite{abeywardana2016layered, gragg2003multi,mitnick2003art,anderson2020security}. 
One study shows that training can reduce social engineering victimization from 62.5\% to 37\% \cite{bullee2015persuasion}.  Training can also help individuals recognize patterns, which can be leveraged to identify social engineering attacks, teach strategies against ongoing attacks, and improve threat appraisal and risk perceptions. 
Training on policies can improve policy compliance \cite{pahnila2007employees, soomro2016information}. Training on strategies against tailgating, shoulder surfing, baiting, and reverse social engineering can reduce such attacks \cite{wang2017coping, jansen2017haisa}. 
    
\smallskip

\noindent{\bf Organizational Policies}. Organizational policies are widely employed defense against social engineering attacks in the physical world \cite{mitnick2003art, cissp2009official}. Organizational policies define expected behaviors and identify information that needs protection \cite{gragg2003multi}. These policies also help reduce uncertainty by defining acceptable practices \cite{greenlees2009iee} and serve as deterrents against specific behaviors \cite{redmon2005mitigation}. For some social engineering attacks, policies may be the only alternative to mitigate them. For example, since dumpster diving is legal (as long as there is no trespassing) \cite{wingo1997dumpster},  establishing corporate policies to define the proper destruction of corporate materials might be the best strategy against dumpster diving. 
However, establishing policies often face a range of challenges. For policies to be enforceable, they must be implemented and monitored. The effectiveness can be affected by multiple factors: one is the culture of an organization subculture \cite{flores2012model,da2017defining};
Another is the attitude in an organization towards compliance and social influence \cite{carmichael2018shrubs}. This is because policy enforcement requires the collaboration of the members of an organization. For example, enforcing a policy targeting tailgating requires individuals to challenge other individuals suspicious of tailgating \cite{redmon2005mitigation} (i.e., exert social influence) and complying with the requirement to display their identification (i.e., display compliance attitude). Simulating Influencing Human Behaviour in Security (SHRUBS) \cite{carmichael2018shrubs} is a tool that examines how psychological aspects and interactions (e.g., beliefs, social norms, the influence of authority figures) can affect the global compliance of a policy. Such an analysis can help identify areas of intervention to improve compliance. Policies can also help minimize the loss incurred by social engineering attacks. For example, a two-factor authentication policy can mitigate the risk when a PIN is compromised via shoulder surfing. 

\smallskip

\noindent{\bf Organizational Procedures}. An organizational procedure provides predefined, step-by-step instructions on addressing a situation, such as strategies for coping with a threat in real-time.  Procedures should be in line with policies and should be part of a training program. In some contexts, procedures are referred to as ``Social Engineering Land Mines'' (SELM) \cite{gragg2003multi}. A SELM is an action that deters an ongoing social engineering attack by surprising the attacker. A ``Justified Know-it-all'' SELM is a person who knows the associated security risks and can handle suspicious events. Other SELM are ``Call-back'' \cite{flores2012model} and ``Please-Hold''  \cite{ghafir2016social} procedures. An extensive list of procedures can be found in \cite{mitnick2003art} under the ``Verification and Authorization Procedure'' section. Procedures can thwart the ``assisting" step in physical reverse social engineering by providing legitimate resources and assistance to targets when they encounter a problem.

\subsubsection{Proactive Defense} 

An effective proactive defense is {\em audit and compliance}. Audit provides an opportunity to measure the effectiveness of a policy and make adjustments to improve the security posture. Audit can help uncover unidentified security weaknesses, which may be caused by poor implementations of policies or the lack of specificity in policies. One study on auditing \cite{cheh2019leveraging} shows that tailgating often occurs when employees escorted visitors into the service rooms.

\section{Characterizing Social Engineering Attack Model, Techniques and Defenses in Cyberspace} 
\label{sec:cyberspace}

Different from the physical world, social engineering attacks in cyberspace are characterized by computer-mediated interactions between a victim and an attacker, which have an effect on the outcome of the interactions (e.g., trust building \cite{riegelsberger2003researcher}).

\subsection{Social Engineering Attack Model in Cyberspace} 
\label{sec:model_se-cy}
As highlighted in Figure \ref{fig:model-in-physical-world}, we consider the following social engineering model of attack in cyberspace.

\subsubsection{Attacker}

The goal and approach of a social engineering attacker in cyberspace are similar to their counterpart in the physical world. However, there are significant differences. First, cyberspace provides an attacker with more resources to personalize attacks, while providing better identity protection mechanisms. Second, the possible anonymity of digital channels allows the attacker to protect their identity while potentially operating in different jurisdictions, which can help with evading legal woes of their actions. Digital channels also allow an attacker to approach multiple victims at once, lowering the cost of waging social engineering attacks and increasing the odds of finding a victim.  As a consequence, social engineering attacks with a low response rate are still profitable \cite{herley2012weis}. 
Third, it is easier to project credibility (for creating trust) in cyberspace than in the physical world because most individuals rely on basic heuristics to judge credibility. Most individuals associate credibility in cyberspace with superficial attributes, such as the professional appearance of a website \cite{dhamija2006acm}, or the presence of high quality and rich content \cite{kim2005dq}. Note that credibility reduces suspicion, improves message persuasion \cite{wathen2002believe}, and increases victims' susceptibility to social engineering attacks \cite{hirsh2012personalized,jakobsson2007privsec,dhamija2006acm}. An attacker's credibility is characterized by the following attributes: \emph{commonality}, \emph{reputation} and \emph{trustworthiness}, which are elaborated below. 

Commonality can be easily established online because  an attacker can use the information in social media and websites to build common ground with a victim.  Information like bias, beliefs, norms, and dialects of a community are useful for an attacker. Using this kind of information, an attacker can impersonate a community member or an acquaintance in online forums or social media groups.  

Reputation is often based on one's network of associates. One method for improving others' perceived reputation of an attacker in cyberspace is to increase the attacker's social media connections with reputable individuals. An attacker can build on a perceived commonality to entice reputable individuals to accept an invitation to connect. Another enticement for attracting reputable individuals in social media is the size of the social network in question.  An attacker can project an extensive social network through the use of bots and fake personas.

Trustworthiness is the perception that the other party is acting in good faith. To project trustworthiness, an attacker can incorporate artifacts (e.g., links, images, graphics) in messages. For example, security indicators like Secure Sockets Layer (SSL)  padlocks \cite{jakobsson2007privsec} or images of organizations which voucher for an individual's trustworthiness, such as Better Business Bureau (BBB) or Federal Deposit Insurance Corporation (FDIC) logos. An attacker can also include URLs that appear to originate from a known trusted site.  An attacker can use a URL that resembles legitimate, well-known URLs (e.g., {\tt www.paypa1.com}, where the letter ``l" is substituted by number ``1" \cite{dhamija2006acm}).  Another approach is to generate a benign-looking malicious URL, such as a long URL which, when shortened by the browser, appears to be benign. For example, an attacker may use {\tt https://myaccount.google.com-securitysettingpage.tk} \cite{heartfield2018protection}.
Finally, an attacker may exploit a victim's trust in a third party, which may be a service or entity trusted by the victim and may provide a communication channel between the victim and the attacker (e.g., dating or employment sites).

\subsubsection{Message} 

As previously mentioned, the purpose of the attacker is to encourage peripheral processing of messages. 
Like in the physical world, an attacker can exploit the same psychological principles to tailor messages, namely: {\em persuasion} \cite{wright2014isr, ferreira2015ieee_st, van2019cognitive}, {\em scamming} \cite{ferreira2015principles}, {\em incentives and motivators} \cite{herley2012weis, cain2018isapp, halevi2015ssr, halevi2013pilot,workman2007iss}, and {\em visceral triggers} \cite{wang2012research}. 

\subsubsection{Victim}

In cyberspace, environment affects a victim's trust and risk perception \cite{riegelsberger2003researcher}. 
Risk perception is an individual's assessment of the risk involving an action, and risk perception affects how much risk an individual is willing to accept \cite{workman2007iss}. 
A victim in cyberspace is also characterized by the following attributes: \emph{domain expertise}, \emph{domain knowledge}, and \emph{vigilance}. 

First, expertise 
in domains other than cybersecurity does not reduce one's susceptibility to social engineering attacks in cyberspace \cite{montanez2020human}. This is because they 
often rely on visual elements and emotions when making decisions involving risks \cite{kumaraguru2006acm}. Additionally, online risk perceptions of non-experts are shaped by the perceived benefit of an activity, and online activities that are considered beneficial are perceived as less risky and performed more often \cite{byrne2016user}. 

Second, applying domain knowledge to recognize social engineering attacks in cyberspace is cognitively demanding because developing domain knowledge often involves learning patterns that indicate malicious intents.  Common pattern identification techniques (e.g., URL parsing) require one to deal with technical complexities, which is cognitively demanding, error-prone and may encourage undue trust \cite{anderson2020security}. This is reasonable because when the risk is high but the situation is difficult to evaluate, trust is an alternative to reduce the complexity of decision-making  \cite{riegelsberger2003researcher}.

Third, vigilance requires attention, which is affected by two components, namely attention {\em switching} and {\em maintenance} \cite{wogalter2018communication}. Attention switching is the process of redirecting attention from one task to another, whereas attention maintenance is the process of dedicating cognitive resources to processing information. Attention switching is a precursor of attention maintenance and leads to central processing \cite{wang2012research}. Salient stimuli trigger attention switching. In ordinary circumstances, an individual's attention is directed to perform the primary task, which supports one's goals (e.g., managing emails, visiting websites, or searching for information online). To detect deception cues in a social engineering message, one must redirect their attention to notice inconsistencies in the message, where the inconsistencies must be salient enough to be detected and trigger the switch. However, a digital environment in cyberspace provides little stimuli to detect deception cues. 
Moreover, the lack of audio and visual cues makes online detection harder \cite{lewis2008cross}. If the deceptive message cues are not salient and suspicion is not triggered, the message is processed via the peripheral route 
and the attacker victimizes a target \cite{vishwanath2011_dssystems}. As a consequence, vigilance is often not triggered, while noting that peripheral route is also affected by individual characteristics (e.g., computer habits \cite{vishwanath2018commresearch}).

\subsection{Social Engineering Attack Techniques in Cyberspace}
\label{sec:cyb-techniques}

As highlighted in Figure \ref{fig:model-in-physical-world}, there are more social engineering attack techniques in cyberspace than the physical world.
We divide social engineering attack techniques in cyberspace into four categories: {\em contextualization}, {\em masquerading}, {\em physical-based access}, and {\em digital reverse engineering}, which are elaborated below.

\begin{figure*}[!htpb]
\centering
\includegraphics[width=\textwidth,keepaspectratio]{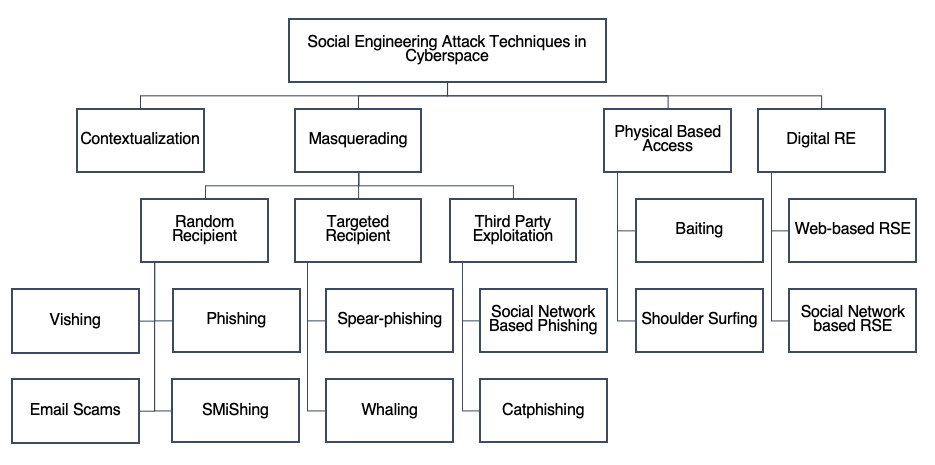}
\caption{Taxonomy of social engineering attack techniques in cyberspace, which are richer than their counterpart in the physical world. 
}
\label{fig:taxonomy-cy}
\end{figure*} 

\ignore{\footnote{\color{magenta} SEKC Techniques under the following tactics are not included: TA-03, TA-05, TA-06, TA-08, TA-09} }

\subsubsection{Contextualization} Contextualization is a message framing technique in which an attacker presents itself as a member of an "in-group" or a community of interest by including details in the message relevant to the group. Examples of details can be current events \cite{alhamar2010ieee,pritom2020characterizing} or topics specific to a community \cite{luo2013cs}. In addition to enhancing the attacker's credibility, contextualization also provides a reason for reducing risk perceptions. Indeed, studies show that contextualization increases social engineering response and compliance rates \cite{goel2017ais, luo2013cs}.  This technique shares some characteristics with the pretexting attack technique in the physical world because it, like pretexting, also provides a false pretense to establish contact.

\subsubsection{Masquerading} 
Masquerading is a family of techniques by which an attacker assumes an online persona to project credibility and trustworthiness. The difference among these techniques lies in the communication channel and the victim selection. Masquerading is similar to impersonation in the physical world and can be further divided into the following three sub-categories: \emph{random recipient}, \emph{targeted recipient}, and \emph{third party exploitation}, where each sub-category has multiple specific techniques.

\smallskip

\noindent{{\bf Random Recipient}}. Attack techniques in this sub-category attempt to contact a large number of potential victims at random. In a crafted message, the attacker usually offers an incentive to motivate the recipients to perform an action. Since the potential victims are selected at random, incentives are based on the general characteristics of the target population. The difference among the  techniques in this sub-category is the transmission medium that is used to carry out the attack, including: email, voice over phone, and text over phone. This sub-category has four specific attack techniques: \emph{phishing}, \emph{SMiShing}, \emph{vishing}, and \emph{email scams}.

First, phishing is the most commonly used attack technique over email. A typical phishing email contains a link that directs a victim to a malicious website for downloading a malicious software (e.g., ransomware) or stealing a victim's credentials. The goal of the message is to motivate the recipient to click the link, for which incentives and urgency cues are widely present, while noting that incentives can vary by groups (e.g., offering free academic products is an effective incentive among college students \cite{abraham2010overview}) and urgency requests are used to urge targets to click a link to avoid losing services \cite{yeboah2014phishing, vishwanath2011_dssystems, alhamar2010ieee}.
            
Second, SMiShing uses text messages or simple messaging service (i.e. SMS) to send social engineering messages. Like phishing, a message includes a link which can cause a target to download malicious content to a mobile device. Unlike phishing, SMS messages have a character count limit, which may force an attacker to leverage psychological elements to trigger an immediate response. For this reason, SMiShing messages include incentive (e.g., explicit content), urgency cues (e.g., an account expiration message), and visceral triggers (e.g., promise of a sexual encounter) to encourage victim compliance \cite{yeboah2014phishing}. The prevalence of SMiShing is due in part to the lack of user awareness of risks and the limited security features available to mobile platform.

Third, vishing (or IVR) uses voice over phone as a medium. It is common in recent years as advancement in Voice-Over-IP (VoIP) allows an attacker to originate a call from any country or area code, regardless of its physical geographical location (i.e., caller ID spoofing) \cite{yeboah2014phishing}.  Caller ID spoofing also allows an attacker to align the message story with observable cues to increase their projected credibility. In addition, the anonymity in the Internet makes tracking the source of a call difficult \cite{griffin2008vishing}. An attacker typically assumes the persona of an authority figure and uses urgency cues to get the victim to release personal information. Since legitimate advertising calls are common, it is difficult for most people to distinguish between vishing and legitimate company calls.

Fourth, email scams attempt to establish relationships with victims on the basis of a common benefit, such as a future financial gain. A well-known attack using this technique is the Nigerian ``419" email scam \cite{stajano2009cambridge, rege2009s}. The crafted message targets the most gullible and susceptible \cite{herley2012weis}.

\smallskip

\noindent{{\bf Targeted Recipient}}. In this sub-category, an attacker tailors or personalizes messages to pre-selected  targets. This sub-category has two specific techniques: \emph{spear-phishing} and \emph{whaling}.
First, spear-phishing personalizes a message based on a distinct characteristic of a victim, such as interest or hobby \cite{alhamar2010ieee}, personality traits \cite{halevi2015ssr} \cite{heartfield2018protection}, or organizational context \cite{pienta2020protecting}. It is typically used when an attacker believes that the victim has access to a desirable asset, such as a company's financial information or transaction systems.  
Second, whaling targets high profile individuals (e.g., C-suite executives, celebrities, wealthy individuals) \cite{koyun2017social}. It requires an attacker to demonstrate intimate knowledge of the victim (e.g., private and public life) as proof of authenticity \cite{pienta2020protecting}. Such information may be acquired by exploiting a victim's network of associates, friends and family members. Whaling is often used on CEO frauds and ghost invoice attacks \cite{junger2020fraud}.

\smallskip

\noindent{{\bf Third-party Trust Exploitation}}. This sub-category of attack techniques targets a specific community hosted by a third-party infrastructure as an attacker can exploit the trust the individuals have of the third party \cite{allodi2019need}. This sub-category includes the following specific attack techniques: \emph{social-network based phishing}, \emph{anglerphising}, and \emph{catphishing}. 
First, social-Network based phishing exploits social media websites to contact potential victims, by creating fake profiles to establish communications with the victims. The resulting attacks are often difficult to detect because there are fewer details in the message about the origin of the message and because users often interpret artifacts like social media connections, friend networks, and profile photos as evidence of credibility \cite{vishwanath2017getting}.  
Second, anglerphishing selects victims based on their social media activities with respect to a brand \cite{o2018angler}. For example, an attacker can send direct messages to victims based on their likes, comments, or tags about a brand. The messages usually contain a link to malicious sites that mirror the brand site. 
Third, catphishing selects victims from online dating sites, by seeking to establish an affection bond and build trust by maintaining frequent communications with a victim  \cite{rege2009s}. Once trust is formed and risk perceptions are lowered, the attacker can request financial assistance. If the victim fails to comply, the attacker uses emotional blackmail or extortion \cite{rege2009s}.

\subsubsection{Physical-based Access} 

This category of attack techniques requires a physical channel to execute, but the objective is to gain access to a network. This category has two specific attack techniques: \emph{baiting} and \emph{shoulder surfing}
First, baiting plants a USB drive containing malicious code at a location that can be easily found (e.g., parking lot or coffee shop \cite{krombholz2015advanced,wagenaar2011usb,salahdine2019social}). When the USB drive is connected to a computer, the malicious code is executed and gives the attacker access to a system of interest. The effectiveness of baiting varies: the success rate for college students varies between 45\% and 98\% \cite{tischer2016users}; the success rate in business setting is about 20\% \cite{mearian2011people}. 
Second, shoulder surfing attempts to gain access to a network by stealing network credentials. An attacker steals a user's credentials (e.g., password or passcode) by watching the user typing them \cite{mitnick2003art, redmon2005mitigation}. The credentials are then used to access the network. Since this attack technique requires an attacker to have physical access to the physical space of a victim \cite{ghafir2016social}, this technique may be executed together with physical impersonations.

\subsubsection{Digital Reverse Social Engineering (RSE)} 
Similar to reverse social engineering in the physical world, this category relies on creating conditions to trigger a victim to reach out to the attacker. The difference is that in digital reverse social engineering, an attacker doesn't have to sabotage a system but simply create the appearance that the system is having problems, while noting that reaching out to the attacker also results in a higher level of trust towards the attacker \cite{irani2011reverse}.
This category has two specific attack techniques: \emph{web-based RSE} and \emph{social network-based RSE}
First, web-based RSE is commonly used to propagate malicious code when a victim encounters a social engineering message and navigates according to the  malicious ads \cite{nelms2016towards, vadrevu2019you}, search result (search engine poisoning) \cite{abraham2010overview}, or a link in webpost (community of interest) \cite{vadrevu2019you}. For example, malicious ads often contain a message which informs the victim of a missing update, application plug-in, or extension to display the content they want to view, which all can include a malicious payload. Legitimate but compromised websites are usually used for malicious ads posting \cite{abraham2010overview}. As another example, search engine poisoning injects malicious websites into the search results.
Second, social network-based RSE exploits social media features to recommend a connection between a victim and an attacker, such as friend recommendations, demographic search, or a victim's page visitor log \cite{irani2011reverse,huber2009towards}. 
For example, demographic search can be exploited to connect a victim to the attacker based on the victim's demographic data.
Social media RSE is less known but effective because individuals are likely to communicate with someone recommended by social media platforms \cite{bilge2009acm}.

\subsection{Social Engineering Defenses in Cyberspace}
\label{sec:defense-cyb}

As described in the methodology, social engineering defenses can be divided into preventive, proactive, active, reactive, and adaptive defenses. Most social engineering defense technologies focus on implementing the latter three defenses, and as a result, they affect human decision-making in similar ways. With an average detection rate of 99\%, these technologies almost eliminate user contact with cyber threats. However, this effectiveness has unintended consequences detrimental to user security performance. When the event rate is less than 1\%, users are more likely to fail to identify social engineering attacks \cite{sawyer2018humanfactors}. Therefore the social engineering messages that avoid detection by the technologies are more likely to be successful. This finding on the effect of low event rate in social engineering detection is similar to the effect of signal detection of low-frequency events and decrements on vigilance \cite{davies1982psychology, proctor2010cumulative}. These social engineering defenses also share approaches to detection. Detection in technology implementation follows one of two approaches: static analysis or dynamic analysis. Static analysis relies on the presence of specific elements in the message (e.g., keywords, URLs, syntax, sender address, etc.) associated with known social engineering attacks. Dynamic analysis relies on statistical analysis and patterns in a message to detect suspicious content. This analysis often involves machine learning and natural language processing to detect suspicious activity. Dynamic analysis social engineering defenses might incorporate static analysis \cite{fette2007learning}, or use existing network logs for cross-validation \cite{ho2017detecting} to improve detection. Finally, the social engineering defense technologies do not require user involvement. 

In what follows, we characterize the social engineering defenses that have been presented in the literature against social engineering attacks in cyberspace, including: {\em preventive defenses}, {\em proactive defenses}, and {\em reactive defenses}, while noting that {\em reactive defenses} do not have their counterpart in the physical world. 

\subsubsection{Preventive Defense}

It would be ideal that social engineering attacks in cyberspace, like other kinds of attack, can be completely prevented because attacks would never succeed. Even though this is not achievable, we should always aim to prevent attacks from succeeding to the extent possible. Preventive defenses against social engineering attacks in cyberspace include: {\em legislation}, {\em access control}, {\em training}, {\em organizational policy}, {\em system and user interface design}, and {\em sandboxing}.

\smallskip

\noindent{\bf Legislation}. In the U.S., laws governing cyberspace are an extension to the laws discussed in Section \ref{sec:phy-prev}. 
The effectiveness of these laws have not been understood because the attribution of social engineering attacks in cyberspace remains to be a hard problem to solve, making it difficult to enforce these laws. Moreover, it is not clear whether the existing laws are adequate in covering all possible kinds of social engineering attacks in cyberspace.

\smallskip

\noindent{\bf Access Control}. Many cybersystems use authentication for access control, and
password is the most common method of authentication. However, password management policies often fail to account for human cognitive limitations. For example, a policy requiring frequent password changes can increase insecure practices, such as generating less secure passwords or reusing passwords \cite{adams1999user}; a policy requiring excessively long and complex passwords can increase memory and cognitive loads.  
An alternative to password is two-factor authentication (2FA), which can be based on  ``something you know and something you have" (e.g., a token and passcode or PIN).
In addition to enhancing authentication, 2FA can help mitigate phishing social engineering attacks \cite{online2018google}. 

\smallskip

\noindent{\bf Training}. 
Training aims to increase individuals' skills for coping with social engineering attacks \cite{pattinson2012imcs, wright2010mis, halevi2016cultural, downs2006acm, arachchilage2014humanbeh, van2017risk}. Training that focuses on providing information on social engineering attacks might not be sufficient because warning users of potential social engineering attacks is not enough. For example, a study \cite{junger2017priming} on the effect of priming and warning on personal information sharing shows that no significant difference between the control group and other groups exposed to priming and warning. 
Another study \cite{jansen2017haisa} shows that exposing individuals to strong fear-appeal messages on phishing increases their protection motivation, which has a direct effect on behavioral intent. Alternatives to in-classroom security education and awareness training are resistance training and security role-play games. Resistance training is the direct exposure of users to a planned social engineering attack, is efficient at quantifying secure behavior instead of secure intent, and can reinforce classroom training and increase security alertness and demonstrate individuals' susceptibility to social engineering attacks \cite{gragg2003multi,flores2013countermeasures}. Finally, games are another alternative to effective training  \cite{irvine2005cyberciege, olanrewaju2015social, newbould2009playing, aladawy2018persuaded}. The advantage of games is that they can expose an individual to a wide array of social engineering attacks (e.g., phishing, pop-up, baiting), levels of sophistication, and various scenarios. Unfortunately, effectiveness of training decreases over time, 
perhaps because of the memorability of training contents and participants' cognitive engagement  \cite{jampen2020don,reinheimer2020investigation}.  


\smallskip

\noindent{\bf Organizational Policies.} In cyberspace, policies can help shape the landscape of an enterprise network and reduce its attack surface. As discussed in Section \ref{sec:phy-prev}, policies can help define acceptable practices and deter unwanted user behaviors. Policies are ineffective when they are counter to human nature  \cite{sasse2001transforming} or organizational culture and norms \cite{flores2013development, da2017defining}. In an organization where information security attitudes are more relaxed, employees might have difficulties adhering to strict security control policies. Some indicators of information security attitudes are information protection, data handling, training, management, security behavior, and compliance. Other factors that affect security policy compliance from a user perspective are incorrect understanding of the applicability of the policy, difficulties in applying policies, or unrealistic and unattainable policies \cite{kirlappos2013comply}. In some organizations, policies that are in line with the dominant individual attitudes can be beneficial. Individual attitudes towards policies are affected by their levels of technical competence --- users with higher technical competence are less likely to favor security policy enhancements  \cite{pfleeger2012leveraging}. In large organizations with a high level of technical competence, delegating responsibility for system updates to users can lead to better attitudes and higher compliance than centralized system management.

\smallskip

\noindent{\bf System and User Interface Design.} In recent years interest in Human-Centered Design and usable security has increased. These approaches can help promote users' secure behaviors by integrating security early in the system design process and communicating security information to enhance secure behaviors.  A common approach to communicating information to the user in applications is to leverage the User Interface (UI), which can be a good mechanism to deliver stimuli to trigger suspicion in users. Although this approach might seem straightforward, developing intuitive UIs that promote security is a challenging task because designers must consider how non-technical users interact with the system and interpret system information \cite{ condori2020can}. Another challenge is to account for temporal psychological attributes that can increase users' susceptibility to social engineering attacks, such as inattentional blindness \cite{pfleeger2012leveraging} and limited cognitive capacity from workload and stress \cite{caputo2014ieee}. 

One of the earlier attempts to leverage UI to communicate security information were browser security toolbars. A browser security toolbar is a browser extension that uses a passive security warning system.  It provides a central location to view all security information for a website. A passive warning alerts the user of the security issues but does not interrupt the user's operations. Multiple studies have demonstrated the inefficiency of this approach \cite{alsharnouby2015phishing, cranor2007phinding, egelman2008you, wu2006security}. Security toolbar inefficiencies owe in part to the lack of understanding of how users make online security decisions. Users determine a website's legitimacy by the appearance of the content in the main content area rather than the information provided in a small area where the security toolbar is displayed \cite{wu2006security}.  In addition to message location, security messages must be salient to grab a user's attention. Toolbar's intuitiveness and level of difficulty for attackers to manipulate also affect its effectiveness \cite{alsharnouby2015phishing}. 

In recent years, the approach to communicating online security information to users has shifted to {\em active warnings}, which differ from {\em passive warnings} in that they display security alert messages in the main content area of the browser, namely the place where users direct most of their attention \cite{alsharnouby2015phishing}.
Additionally, interrupting users from performing their primary task forces them to switch their attention to the warning. Active warnings have shown to be more effective in promoting secure behavior than passive warnings. One study \cite{egelman2008you} shows that 79\% of the participants receiving an active warning message follow the advice, which is in sharp contrast to the 13\% when using a passive warning message.

\smallskip

\noindent{\bf Sandboxing.} A sandbox is an isolated environment inside another computing environment. It creates a layer of isolation between the main computing environment and the application running in the sandbox, effectively preventing a malicious application running in the sandbox from gaining access to a system's resources outside the sandbox. Sandbox can help prevent social engineering attacks. For example, in web browsers, sandboxes can prevent the manipulation of visual components of a website. In social engineering attacks that involve the download of malicious code, sandboxes can also help mitigate them \cite{heartfield2018protection, heartfield2018detecting}.



\subsubsection{Proactive Defenses}

We classify these defenses into two categories: {\em audit and compliance} and {\em security (threat) intelligence}.

\smallskip

\noindent{\bf Audit and Compliance.} Auditing is a mechanism to test the effectiveness of policies and can offer insights into users' secure behaviors that cannot be captured otherwise.  One study on testing the effectiveness of security policies \cite{orgill2004urgency} shows that participants willingly provide their usernames (81\%) and passwords (59\%) to the auditor, who is able to acquire an all-day access card to the entire facility and access sensitive information (e.g., credit cards, copy of a master key). Penetration tests can supplement audits. When an audit involves a penetration test, it is important to consider the psychological toll that can take on victimized individuals. Several options are available to minimize the possible psychological toll on the victims \cite{jakobssondesigning, dimkov2010csac}.

\smallskip

\noindent{\bf Security (Threat) Intelligence.} TSecurity intelligence is the sharing of information about threats and actors to prevent or mitigate an attack. Security intelligence can be generated by different sources. Currently, most of its use focuses on cyber investigations. However, it can also be used for proactive defense.  One approach to generate security intelligence is through crowd-sourcing  \cite{vincent2019don}.  In this approach, the users share information about malicious social engineering attack encounters to warn others. Sharing information about a recent or ongoing social engineering attack with the users has the potential of increasing user's vigilance and protection attitudes \cite{jansen2017haisa}. Security Intelligence is most beneficial when it also includes a coping strategy to protect against the threat. 

\ignore{

\cite{ford2009ieee} automating the analysis of malicious ads that use Flash. Based on characteristics extracted from analysis of malicious ads. Ads are identified as malicious based on their behavior or their characteristics.

\cite{jayasinghe2014efficient} Java Script drive-by downloads. Dynamics allows the analysis of at run time deobfuscated code. 
}

\subsubsection{Reactive Defenses}

We classify these defenses into two categories: {\em network-based detection}, {\em behavior-based detection}, and {\em real-time detection}. 

\smallskip

\noindent{\bf Network-based Detection.} Most social engineering defense technologies identify patterns associated with known social engineering attacks. In most networks, social engineering defense technologies are layered \cite{flores2013countermeasures}, and each technology follows a different detection approach. One specific approach is to combine static and dynamic analysis to detect social engineering attacks (e.g., emails). 
Artificial Intelligence / Machine Learning (AI/ML) has been widely used to detect such attacks (see, e.g., \cite{gutierrez2018learning,XuCodaspy13-maliciousURL,XuCNS2014,hamid2011phishing,pritom2020characterizing,MirIEEEISI2020MalciousWebsites}).
Cognitive features, like cognitive vulnerability triggers, can be leveraged to detect and prioritize suspicious emails based on their likelihood of success \cite{van2019cognitive}. The study shows that in a financial institution, successful phishing emails contain more cognitive vulnerability triggers with a message topic relevant to the members of the organization.

\smallskip

\noindent{\bf Behavior-based Detection.}  This defense leverages behavioral patterns of entities (e.g., application, system, or user) over an extended period to detect attacks \cite{stringhini2015ain}. 
For example, user email behaviors can be derived from previous emails and user email habits, including writing style (e.g., word patterns, use of unique words), composition style (e.g., email activity patterns, recurrent URLs in the emails), and social patterns (e.g., common email interactions, personal address list).  
This defense can be effective against CEO frauds \cite{junger2020fraud}, which use compromised email accounts of executives to request the execution of an unauthorized financial transaction. This is because an attacker's email behavior would be different from an executive's.
This approach can also be used to detect fake social media entities, for example, by using honeypots to identify social media spammers \cite{lee2010acm}.
Fake media entities can also be detected using the mathematical properties of their social graph, while leveraging social psychological factors (e.g., persuasion mechanism, enticement) in modeling the social graph  \cite{alvisi2013sok}. 

\smallskip

\noindent{\bf Real-time Detection.} In web-based social engineering attacks, dynamic rendering of webpage content is often leveraged to avoid detection via various kinds of evasion techniques (e.g., code obfuscation or behavior modification based on the platform), while leveraging AI/ML techniques \cite{XuCodaspy13-maliciousURL,XuCNS2014}.  Dynamic rendering content changes on each session based on user properties like geographical region, language, or browsing history.  
Digital reverse social engineering (RSE) attacks use dynamic rendering content. Examples of RSE vectors that use dynamic rendering are search results, Web advertisements, and web posts.  In addition to facilitating evasion, dynamic content in RSE can tailor attacks based on victims' preferences and leverage attention-grabbing elements \cite{nelms2016towards}. 

\ignore{
\footnote{the subsection structure may need to be refined; once refined/finalized, putting the content of the subsequent section into this one}

\subsubsection{Which Social Engineering Attacks in Cyberspace Have Been Adequately Defended?}

maybe none?

\subsubsection{Which Social Engineering Attacks in Cyberspace Have Not Been Adequately Defended?}

}

\section{Contrast Analysis}
\label{sec:contrast-analysis}

Our contrast analysis is driven by the following question: (i) What are common to, and different between, social engineering attacks in the physical world and in cyberspace?
(ii) Which defenses designed for the physical world can be adapted to defenses for cyberspace, and vice versa?
(iii) What defenses must be specifically designed to cope with social engineering attacks in cyberspace because they are unique to this domain?

\subsection{Social Engineering Attacks in the Physical World vs. Cyberspace}

Social engineering attacks in the physical world and their counterpart in cyberspace have much in common. First, they both use messages containing persuasive content, visceral triggers, and deception elements. Second, they exploit the same psychological attributes to conduct attacker-victim interactions. For example, the perceived credibility of an attacker is crucial to the success of the attack, whereas triggering individuals' suspicion is key to achieve effective defense. Third, humans are often blamed for system failures. However, the lack of human perspective in the design results in solutions that are incoherent with human cognitive processes, and that introduces weakness in the system after its release \cite{leveson1995safeware} . 

\begin{insight}
Future system designs should strive to achieve resilience against social engineering attacks or {\em social engineering resistance by design}. 
\end{insight}

There are significant differences between them. First, unlike social engineering attacks in the physical world, persuasive content used in cyberspace can be intercepted and analyzed for linguistic elements before it reaches the user. This mediation offers opportunities for designing new defenses in cyberspace based on message persuasiveness.  Second, there is a difference in how compliance is affected by the environment: achieving compliance in the physical world can be aided through visual elements and interactions, but compliance in cyberspace is achieved through establishing relationships with a victim.  Third, effective defenses in the physical world and cyberspace would be quite different because domain knowledge significantly reduces scam victimization in the physical world but this does not hold true in cyberspace.  Fourth, social engineering attacks in cyberspace are not restricted to geographical boundaries, which makes it difficult to design preventive defenses such as legislation.

\begin{insight}
While social engineering attacks in the physical world and their counterpart in cyberspace have much in common, there are significant differences. This means that we need to design effective defenses in both worlds, respectively.
\end{insight}

\subsection{Adapting Defenses from One World to the Other}

Discussions in Sections \ref{sec:defense-phy} and \ref{sec:defense-cyb} suggest that the following defenses in the physical world may be adapted to cyberspace. First, the physical world provides mechanisms to verify the credibility of an individual or organization, partly owing to legislative requirements. Although these mechanisms have limitations, they have contributed to reducing the prevalence of social engineering attacks in the physical world. This has implications for designing defenses in cyberspace, where a credibility verification process must safeguard against attacker manipulations. Moreover, defenses in cyberspace must also be compatible with human cognitive models of credibility in the physical world. Second, known social engineering attacks in the physical world could be leveraged to derive threat detection patterns in cyberspace. For example, the ability to reach a wider audience and the low cost of operations facilitate the migration of criminal enterprises from the physical world to cyberspace \cite{langenderfer2001consumer}. Evidence of this migration is the variance of the recruiting scams \cite{stajano2009cambridge,10.1002/spe.2180,
allodi2019need}. This situation presents a unique opportunity to build resilient cybersystems by leveraging the knowledge of social engineering attacks in the physical world. 

\begin{insight}
The strategies and processes that are designed to prevent social engineering attacks in the physical world can be leveraged to design effective defenses in cyberspace that are compatible with human cognition models.
\end{insight}

Discussions in Sections \ref{sec:defense-phy} and \ref{sec:defense-cyb} also suggest that the following defenses in cyberspace may be adapted to the physical world. First, the human-centered design approach in cyberspace can bring new insights into our understanding of human information processing from a security and protection perspective. Some of these insights might be transferable to the physical world to gain a deeper understanding of crime prevention. Second, existing detection mechanisms in cyberspace might help proactively warn of potential scams and frauds in the physical world. As criminal enterprises operate across the physical world and cyberspace, detection in cyberspace could serve as an early-warning mechanism for similar attacks in the physical world. For example, knowing that social engineering attackers are leveraging significant events occurring in the physical world \cite{MirIEEEISI2020MalciousWebsites}, defenders can monitor for keywords associated with these events to actively detect new social engineering attacks.

\begin{insight}
Reactive defense in cyberspace might be help identify attack patterns and provide early-warning for social engineering attack in the physical world.
\end{insight}

\subsection{Designing Defenses in Cyberspace}

Since our focus is on defending against social engineering attacks in cyberspace, our study suggesting the following directions for future research. 
First, we need to understand the root causes to human susceptibility to social engineering attacks in cyberspace. This is critical to design effective defenses. Second, there is a lack of stimuli in cyberspace to detect deception cues and trigger vigilance. Therefore, it is important to investigate how to help users detect deception cues and trigger their vigilance. For this purpose, it is critical to understand how online information is processed by humans and how to trigger an individual's protective response online.
Third, how can we develop effective domain knowledge to effectively train individuals to resist social engineering attacks in cyberspace? Self-efficacy (i.e., one's ability to address a situation) plays an important role in protection motivation, but a user's confidence in handling a threat does not equate to social engineering attack outcomes. The causes of these deviations are not well understood. An answer could provide an insight into how to promote online protection behavior. 

\begin{insight}
There are fundamental problems in understanding and addressing social engineering attacks in cyberspace that remain largely open.
\end{insight}

\section{Conclusion}
\label{sec:conclusion}

We have presented a unified terminology and methodology for characterizing and understanding social engineering attacks in the physical world and cyberspace. Grounded in individual and social cognition, the methodology included a novel model for describing social engineering attacker-victim interactions, including the psychological factors that are relevant to social engineering attacks. The methodology also guided us to systematize social engineering attacks and defenses in the physical world and cyberspace. In particular, our contrast analysis of the social engineering attacks and defenses in the physical world and cyberspace led to a number of insights, which shed light on future research. As shown, our understanding of social engineering attacks, especially in cyberspace, is superficial as fundamental problems are yet to be answered.

\smallskip

\noindent{\bf Acknowledgment}.
The first author is also affiliated  with The MITRE Corporation, which is provided for identification purposes only and is not intended to convey or imply MITRE's concurrence with, or support for, the positions, opinions, or viewpoints expressed by the author.  This work was supported in part by ARO Grant \#W911NF-17-1-0566, NSF Grants \#2122631 (\#1814825) and \#2115134, and Colorado State Bill 18-086.

Approved for Public Release; Distribution Unlimited. Public Release Case Number 21-2666. ©2021 The MITRE Corporation. ALL RIGHTS RESERVED.

\bibliography{main,metrics1}

\end{document}